\documentclass[aps,amsfonts,amsmath,superscriptaddress,amssymb,prb,twocolumn]{revtex4}

\usepackage{graphicx}
\usepackage{dcolumn}
\usepackage{bm}
\usepackage{longtable}

\begin{document}

\title{DISTINCT OXYGEN HOLE DOPING IN DIFFERENT LAYERS OF $\rm Sr_{2}CuO_{4-\delta}-La_{2}CuO_4$ SUPERLATTICES}

\author{S. Smadici}
\affiliation{Frederick Seitz Materials Research Laboratory, University of Illinois, Urbana, IL 61801, USA}%

\author{J.C.T. Lee}
\affiliation{Frederick Seitz Materials Research Laboratory, University of Illinois, Urbana, IL 61801, USA}%

\author{P. Abbamonte}
\affiliation{Frederick Seitz Materials Research Laboratory, University of Illinois, Urbana, IL 61801, USA}%

\author{A. Rusydi}
\affiliation{NUSSNI-NanoCore, Department of Physics, National University of Singapore, 117542, Singapore}%

\author{G. Logvenov}
\thanks{Present address: Max-Planck-Institut for Solid State Research, Heisenberg Strasse 1, D-70569, Stuttgart, Germany.}
\affiliation{Brookhaven National Laboratory, Upton, NY 11973, USA}%

\author{I. Bozovic}
\affiliation{Brookhaven National Laboratory, Upton, NY 11973, USA}%

\begin{abstract}

X-ray absorption in $\rm Sr_{2}CuO_{4-\delta}-La_{2}CuO_4$ (SCO-LCO)
superlattices shows a variable occupation with doping of a hole
state different from holes doped for $x \lesssim x_{optimal}$ in
bulk $\rm La_{2-x}Sr_{x}CuO_4$ and suggests that this hole state is
on apical oxygen atoms and polarized in the $a-b$ plane. Considering
the surface reflectivity gives a good qualitative description of the
line shapes of resonant soft X-ray scattering. The interference
between superlattice and surface reflections was used to distinguish
between scatterers in the SCO and the LCO layers, with the two hole
states maximized in different layers of the superlattice.

\end{abstract}

\maketitle

\section{Introduction}

Hole doping of $\rm La_{2-x}Sr_{x}CuO_{4}$ is described in the
Zhang-Rice singlet (ZRS)-upper Hubbard band (UHB) model for
$0<x\lesssim 0.2$ as ZRS states on in-plane oxygen atoms. These
holes are visible as a feature in X-ray absorption spectroscopy
(XAS) at the O K edge, called the mobile carrier peak (MCP). The MCP
gains intensity with doping at the expense of LHB (lower Hubbard
band) states and concurrent ``spectral weight transfer" from UHB
states. This model, expanding the Cu $d_{x^{2}-y^{2}}$ one-band
Mott-Hubbard model to the Cu $d_{x^{2}-y^{2}}$, O $p_x$, O $p_y$
three bands of the $\rm CuO_{2}$ planes, describes well the
variation, observed with XAS at the O K edge, of unoccupied density
of states in cuprates for relatively low doping ($x\lesssim
0.2$)~\cite{1991Eskes,1991Chen}.

However, the variation of the maximum critical temperature $\rm
T_{c,max}$ between different superconducting compounds cannot be
explained within the UHB-ZRS model and its in-plane orbitals only
and possible extensions of the model to out-of-plane orbitals have
been intensively investigated. Since there can be only one Fermi
surface for an isolated $\rm CuO_2$ plane, other indications of the
relevance of out-of-plane orbitals came from angle-resolved
photoemission (ARPES) measurements of square- and diamond-shaped
Fermi surfaces for $\rm La_{2}CuO_4$ (Ref. 3) and $\rm
Ca_{2}CuO_{2}Cl_{2}$ (Ref. 4). The most relevant out-of-plane
orbital that hybridizes with states in the $\rm CuO_{2}$ plane is
the apical oxygen $p_{z}$ orbital mixed with the Cu
$3d_{3z^{2}-r^{2}}$ (Refs. 5,6) or $4s$ (Ref. 7) orbital in the $\rm
CuO_{2}$ planes. The occupation of the apical oxygen orbitals
modifies the bond valence sums and was used to explain general
trends of $\rm T_{c,max}$.~\cite{1990deLeeuw, 1991Ohta} The effect
of apical oxygen $p_z$ energy level on parameters of an expanded
$t-J$ model has been considered for different
materials.~\cite{2009Yin} In contrast, the axial hybrid between the
O $p_z$ and Cu $4s$ orbitals, with $\rm Cu$
$d_{3z^{2}-r^{2}}$-states occupied, has been addressed in Ref. 7.
These calculations predicted that an empty apical oxygen $p_z$
orbital modifies the in-plane hole hopping parameter $t'$ between
sites along the orthorhombic axes, consistent with ARPES
measurements, with a suppressed $t'$ from the presence of unoccupied
apical oxygen orbitals correlating with a smaller $\rm
T_{c,max}$.~\cite{2009Yin, 2001Pavarini, 1996Raimondi}

In addition, the depletion of the UHB states in bulk $\rm
La_{2-x}Sr_{x}CuO_{4}$ at $x\sim 0.2$,~\cite{1991Chen} shows that
the ZRS-UHB model also needs to be modified to describe the hole
doping for higher $x$. However, because of limits of bulk crystal
growth, doping dependence studies have been limited to doping near
$x=2$, realized in bulk $\rm Sr_{2}CuO_{4-\delta}$ with a very large
$\rm
T_c$,~\cite{2006Yang,1994Al-Mamouri,1999Choy,1994Shimakawa,1995Zhang,2007Yang,2009Geballe}
where the oxygen atoms are removed from the structure ($\delta >
0$), and for relatively low $x$, near the superconducting dome,
where there is evidence of a qualitative change with $x$ in the
doping process near $x_{optimal}$. Specifically, the effective Cu
ion magnetic moment in $\rm La_{2-x}Sr_{x}CuO_4$ and the magnetic
exchange between Cu spins are strongly reduced~\cite{1989Johnston}
at $x\sim 0.2$ and the $T$-independent Pauli paramagnetism is
replaced at $x\sim 0.22$ by $T$-dependent Curie paramagnetism with
increased doping.~\cite{2005Wakimoto}
Calculations~\cite{1991DiCastro, 1992Feiner} suggested that the
doping mechanism is different for $x
> x_{optimal}$, as the number of $a_{1}$-symmetry states, related to
the Cu $3d_{3z^{2}-r^{2}}$ orbital~\cite{2009Yin} and exceeding the
optimal doping $x_{optimal}$, correlated with $\rm T_{c,max}$. A
small but growing contribution from apical oxygen $p_z$ orbital
holes mixed with Cu $d_{3z^{2}-r^{2}}$ orbitals at MCP was inferred
from angle-resolved XAS measurements~\cite{1992Chen} and from the
variation of the Cu-apical oxygen distance with
doping~\cite{2005Bozin}, which also correlates with a variation in
$\rm T_{c}$.~\cite{2009Butko,2003Karimoto} Recently, a saturation of
the XAS MCP intensity with increasing doping near $x_{optimal}$ was
observed~\cite{2009Peets} and dynamical mean-field theory
calculations~\cite{2010Millis} concluded that either additional
orbitals become relevant in this doping range or that new model
parameters would be needed to account for multiple-hole
interactions. The additional orbital or band to consider in the $\rm
La_{2-x}Sr_{x}CuO_{4}$ hole doping process between $x \sim 0.2$ and
$x \sim 2$ is not known.

\begin{figure}
\centering\rotatebox{0}{\includegraphics[scale=0.65]{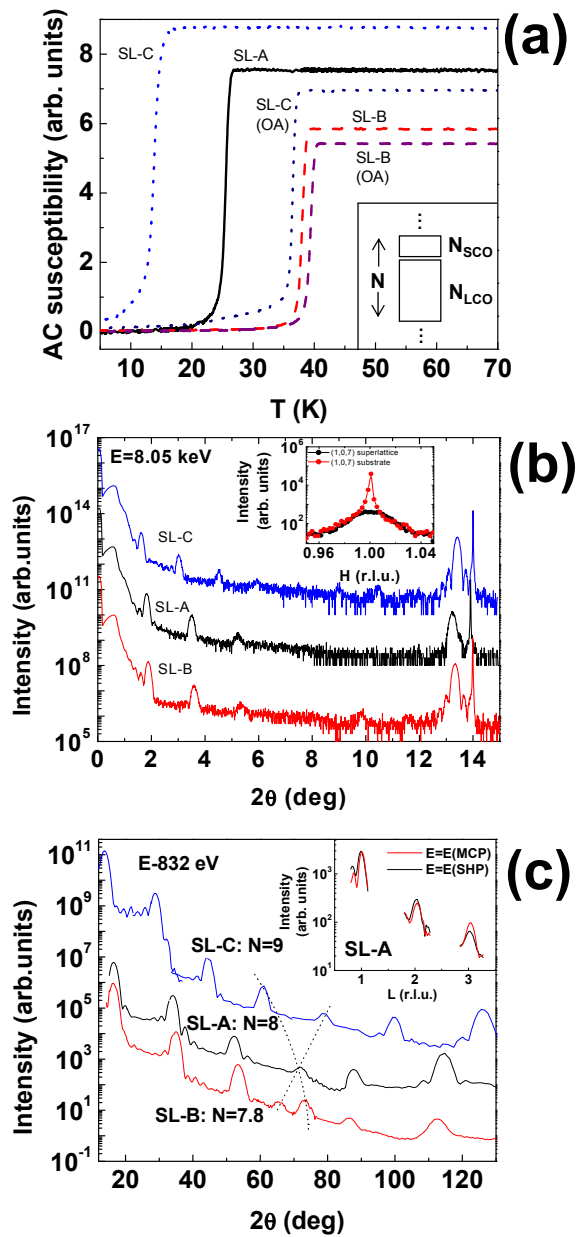}}
\caption{\label{fig:Figure1} (Color online) (a) Measurements before
and after ozone annealing (OA) show that all SL are superconducting.
The inset shows a sketch of one SL superperiod. The atomic planes
within the SCO and LCO layers are shown in Fig. 6(a). (b) Hard X-ray
reflectivity characterization ($E=8.05~\rm keV$). The sharp peak at
$2\theta \approx 14^{\circ}$ is the substrate (002) reflection peak,
with $\rm c_{LSAO}=12.64~\AA$.~\cite{2009Butko} Using this peak as a
reference marker, HXD measurements showed that the average ML
thickness was $\rm 6.64~\AA$, $\rm 6.63~\AA$ and $\rm 6.59~\AA$ for
SL-A, SL-B and SL-C, respectively. The difference is consistent with
the thicknesses of the SCO and LCO layers in each SL and the larger
$c$-axis parameter of SCO films $\rm
c_{SCO}=13.55~\AA$~\cite{2003Karimoto} compared to that of LCO films
$\rm c_{LCO}=13.3~\AA$. Superlattices with smaller LCO layer
thickness in a superperiod (not shown) had less good growth. The
inset shows the epitaxial growth, where $\rm
a_{LSAO}=3.76~\AA$.~\cite{2009Butko} (c) Characterization with soft
X-rays near the La edge ($E=832~\rm eV$). The dotted lines show the
evolution of the SL peaks with superperiod
thickness.~\cite{1942Hendricks} Inset shows L scans at MCP and SHP
energies (Sec. II D). The difference at L=3 between MCP and SHP
amplitudes is also visible in Fig. 4.}
\end{figure}

In this study, doping in the range $1 \lesssim x \lesssim 1.5$,
inaccessible with bulk crystal growth techniques, was obtained with
$\rm Sr_{2}CuO_{4-\delta}-La_{2}CuO_4$ (SCO-LCO) superlattice (SL)
growth. We observed that occupations of two distinct oxygen hole
states are gradually modified with doping. Using the interference of
the SL reflections with the surface reflection, we determined that
the two hole states are maximized in different layers. This suggests
that the states emptied preferentially in the SCO layers are the
additional states in the extension of the ZRS-UHB model to this
doping range.

\section{Experiments}

\subsection{Superlattice structure}

The superlattice samples were grown by molecular beam epitaxy on
$\rm LaSrAlO_4$ (LSAO) substrates at Brookhaven National Laboratory.
Three SL, called SL-A, SL-B and SL-C in the following, were chosen
after atomic force microscopy (AFM) and hard X-ray diffraction (HXD)
measurements. HXD measurements [Fig. 1(b)] were made using a Philips
X' Pert diffractometer. From a Nelson-Riley fitting, the number of
layers in one superperiod for SL-A and SL-C is $N=7.81\pm 0.1~\rm
ML$ and $N=9.27\pm 0.3~\rm ML$ respectively, where $\rm
1~ML~(``molecular ~layer")$ is the average d-spacing of one SL
``formula-unit" layer (half the unit cell), given by the (002)
reflection. The SL structure was further characterized with resonant
soft X-ray scattering (RSXS) measurements at beamline X1B at the
National Synchrotron Light Source. Reflectivity measurements near
the La edge [Fig. 1(c)] for $Q=(0,0,Q_{z})$, where $Q_{z}=2\pi
L/c_{SL}$ is the scattering momentum in units of SL superperiod
$c_{SL}$, show superperiods with an integer number of layers,
$N=8~\rm ML$ and $N=9~\rm ML$ for SL-A and SL-C respectively,
consistent with the HXD measurements. The slightly worse SL-B has
$N=7.8~\rm ML$. The thickness of the SCO layers within a SL
superperiod is $N_{SCO}\approx 2~\rm ML$ for all samples, but
somewhat larger for SL-A compared to SL-B or SL-C, because of larger
doping (Sec. II B). The number of repeats was 8, 7 and 8 for SL-A,
SL-B and SL-C, respectively. A sketch of one SL superperiod is shown
in Fig. 1(a) (inset).

The superconducting critical temperatures from AC susceptibility
measurements were 25 K for SL-A, 38/39.5 K for for SL-B and 14/36.5
K for SL-C, where the first and second values for SL-B and SL-C are
for measurements before and after ozone annealing at $\rm
350~^{\circ} C$ for $20$ min [Fig. 1(a)]. AFM images were taken on a
Dimension 3100 instrument. They showed a surface covered by islands
approximately 150 nm wide and 2 nm high for SL-B (data not shown);
SL-A or SL-C did not have these features. The AFM surface RMS
roughnesses $\sigma_{s}$ for SL-A, SL-B and SL-C were 0.61 nm, 0.76
nm, and 0.43 nm, respectively.

\begin{figure}
\centering\rotatebox{0}{\includegraphics[scale=0.55]{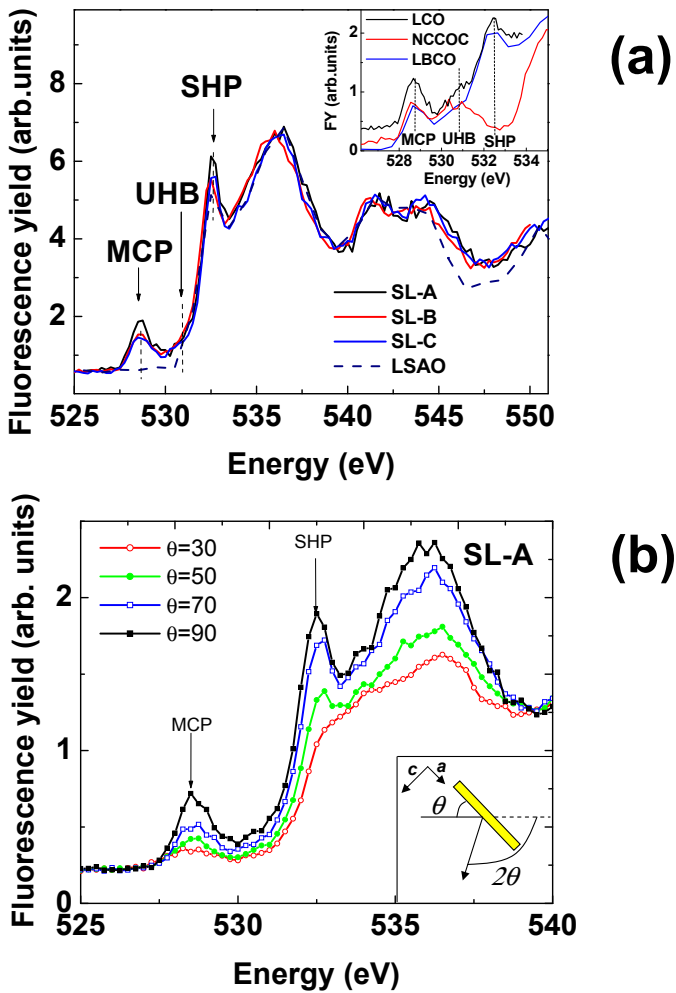}}
\caption{\label{fig:Figure3} (Color online) (a) Superlattice
fluorescence yields differ at the MCP and SHP energies only. The
relative intensities of the MCP peaks in fluorescence yield make it
clear that SL-A has more Sr than SL-B and SL-C. Since the LSAO
substrate SHP is smaller than that of either SL, sampling more
substrate FY in SL-A measurements (SL-A is thinner than SL-C) cannot
explain the higher SHP intensity. The inset shows for comparison FY
for bulk $\rm La_{2-x}Ba_{x}CuO_4$ (with $x=0.125$), $\rm
La_{2}CuO_{4+\delta}$ (with $\delta=0.12$) and $\rm
Na_{x}Ca_{2-x}CuO_{2}Cl_{2}$ (with $x=0.08$). (b) Angle-resolved
fluorescence yield on SL-A. The measurements have been aligned at
525 eV and 560 eV. Inset shows the measurement geometry. The
detector angle $2\theta$ was kept fixed at $110^{\circ}$.}
\end{figure}

The interface roughness can be characterized with HXD reflectivity.
The $L=1$ reflection width [Fig. 1(b)], dominated by the total SL
thickness and not sufficiently sensitive to small-scale roughness,
was approximately the same for all SL. However, the $L=2$ linewidth
was the same for SL-A and SL-C and $9~\%$ higher for SL-B. Also, the
$I(L=2)/I(L=1)$ ratio was the same for SL-A and SL-C and $23~\%$
smaller for SL-B. These observations suggest that SL-A or SL-C have
a smaller interface roughness than SL-B, consistent with the AFM and
superperiod measurements. In addition, although the surface
roughness $\sigma_{s}$ of SL-A is larger than that of SL-C, their
interface roughnesses $\sigma_{i}$ are similar. This will be used to
explain the difference in scattering at the O edge between SL-A and
SL-C (Sec. III B). For an estimate of the $\rm Sr$ doping range we
use a SL interface RMS roughness $\sigma_{i}\sim \rm 6~\AA$. The
maximum Sr doping is estimated with either a ``flat-top" or a
Poisson distribution as $x_{max} \sim 1.5$ and $x_{max} \sim 1$ for
a SL with 2 ML and 3 ML thick SCO layers, respectively, outside the
current possibilities of bulk crystal growth. Therefore, the middle
nominally SCO layer in a 3 ML thick SCO layer is approximately $\rm
La_{0.5}Sr_{1.5}CuO_{4}$ and the two nominally SCO layers in a 2 ML
thick SCO layer are approximately $\rm LaSrCuO_{4}$. The shorthand
notation ``SCO" and ``LCO" from the deposition sequence will
continue to be used for simplicity for the SL layers. Our
conclusions do not depend on the exact values of SCO and LCO layer
thickness or interface roughness.

\subsection{Oxygen edge fluorescence yield}

The first indication of hole doping at two distinct energies comes
from XAS measurements with fluorescence yield (FY) detection and
$\pi$-polarized incident light, made at beamline X1B at the National
Synchrotron Light Source. The sample and detector angles were
$\theta = 80^{\circ}$ and $2\theta=110^{\circ}$ respectively,
defined as shown in Fig. 2(b) (inset). SL measurements are shown in
Fig. 2(a). FY of bulk $\rm La_{2-x}Sr_{x}CuO_{4}$ (LSCO) has three
main low-energy features~\cite{1991Chen}: the MCP of the ZRS state,
the UHB feature and a peak called here the ``second hole peak"
(SHP). The highest peak at $\rm 536.1~eV$, sometimes associated with
orbitals mixed with La states~\cite{1991Kuiper}, is followed by
``continuum oscillations". There is no clearly discernible UHB peak
in SL FY. The UHB intensity in bulk LSCO is
negligible~\cite{1991Chen} for $x> 0.15$; therefore, the nominally
undoped LCO layers in the SL are instead thoroughly doped with MCP
holes (the occupation factor $\langle t_{l,LCO}^{MCP}\rangle
>0.15$), consistent with estimates based on roughness and hole
diffusion length~\cite{2009S}.

\begin{figure}
\centering\rotatebox{0}{\includegraphics[scale=0.75]{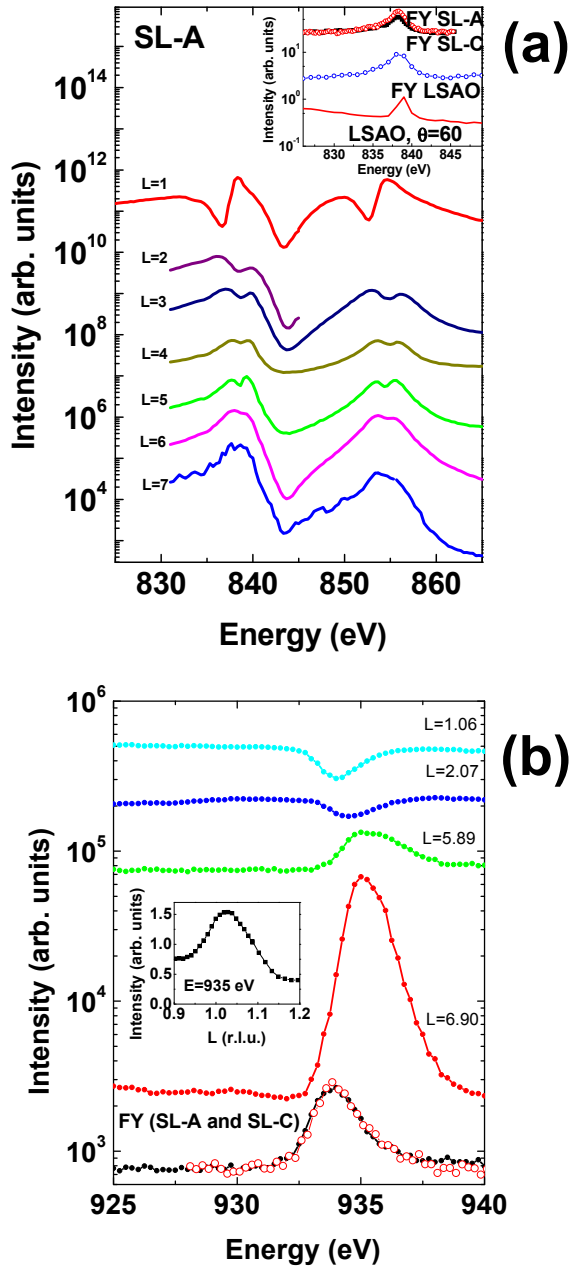}}
\caption{\label{fig:Figure4} (Color online) (a) Reflectivity at the
La edge for SL-A. The plots have been shifted vertically for
clarity. Inset: FY for SL-A and SL-C. The only clear difference in
FY is an increase in the overall height, consistent with a larger
number of La atoms in SL-C. The two lower curves are the LSAO
substrate FY and reflectivity at $\theta=60^{\circ}$, showing the
strong resonant increase in the substrate reflectivity. The
difference from measurements on SL is due to the absence of SL
scattering contrast $\delta f_{SL}$ (Sec. III A). (b) Line shapes at
the Cu edge. For intermediate conditions the scattering line shape
is flat: this is observed for SL-C at L=4 (not shown). The
two-dimensional profile has a saddle point for L=1: it is a peak
when scanning $L$ (inset) and a dip when scanning $E$. This is in
contrast to multiple-scattering effects in HXD that can be observed
for certain oxide SL, where a dip is present at integer $L$ instead
of a peak (data not shown). The lower curves compare the FY for SL-A
(full symbols) and SL-C (open circles).}
\end{figure}

The SL FY spectra in Fig. 2(a) have been aligned below and above the
edge. The alignment normalizes out the variation in the number of
oxygen atoms and therefore, the remaining differences between FY
spectra are due to electronic contrast from the difference in
valences of oxygen atoms. For SL-A and SL-C this occurs exclusively
at MCP and SHP energies. Therefore, the SHP peak is related to
doping of oxygen states, not to a structural defect, e. g. a vacancy
or interstitial oxygen. Further support for this interpretation
comes from FY measurements of bulk LSAO, where the SHP peak is
observed (Fig. 2); since LSAO has no vacancies or interstitial
atoms, SHP cannot be related to these defects. This is also
consistent with the observation that SL-A has a larger SHP intensity
than SL-C, while compounds without apical oxygen atoms, e. g. NCCOC,
have a smaller intensity at SHP [Fig. 2(a), inset]. These FY results
are a clear indication that the ZRS-UHB model extension in our
doping range beyond the MCP and UHB levels is related to SHP and not
to the vacancies of bulk SCO.

The SHP energy is relatively insensitive to the type of oxygen
neighbors. For instance, it is not influenced by the Cu $d$ states
since $\rm Al^{3+}$ in LSAO has no $d$ states or by replacing La
with Sr. The FY increase from SL-C to SL-A is at the same energy
(SHP) as the energy of a peak of the undoped LCO compound, which
suggests that the SHP state, partly empty in undoped LCO, is
gradually emptied further, rather then removed, with additional hole
doping in SL-A compared to SL-C; that is, doping also makes holes at
SHP. Therefore, the total numbers of MCP and SHP holes in one
superperiod are related by the charge conservation equation as
$\sum_{l}\langle t_{l}^{MCP}\rangle+\sum_{l} \langle
t_{l}^{SHP}\rangle=2N_{\rm SCO}$, where $l$ sums over one
superperiod, $N_{\rm SCO}$ is the number of SCO layers and $2N_{\rm
SCO}$ is the total number of doped holes in one superperiod (each
$\rm Sr_{2}CuO_{4-\delta}$ layer dopes 2 holes when $\delta \approx
0$).

Knowledge of the SCO oxygen vacancy site would suggest the site of
SHP; however, the question of the vacancy site is not
settled.~\cite{2009Geballe} The absence of this feature in the
cuprate NCCOC, which does not have the apical oxygen (Fig. 2),
suggests that SHP is an apical oxygen state. It is possible that SHP
is absent in NCCOC for other reasons than the absence of the apical
oxygen. For instance, in an alternative view of angle-resolved FY
measurements on bulk LNO~\cite{1991Kuiper,1998Kuiper}, SHP are
considered in-plane oxygen states mixed with Ni, and polarized
in-plane. However, the SHP energy is not influenced by the Cu $d$
states or replacing La with Sr, which suggests various
hybridizations are not essential. The incipient out-of-plane doping
for low $x$ (Refs. 2, 22) supports the view that doping is not
confined to the $\rm CuO_{2}$ planes for high $x$. Indeed, more
detailed early calculations of the $\rm La_{2-x}Sr_{x}CuO_4$
system~\cite{1988Guo} suggested that the shift with doping in the
relative alignment of the in-plane [O(1)] oxygen and apical [O(2)]
oxygen ionization potentials will eventually lead to the
preferential emptying of the apical oxygen orbitals at larger $x$.
If SHP is an apical state, the Cu ions would be surrounded by holes
on all six neighboring oxygen atoms. This transition to a gradual
emptying of apical oxygen orbitals, balancing the valences of all
six oxygen atoms near a Cu site (this hole distribution is seen in
undoped $\rm La_{2}NiO_{4}$~\cite{1991Kuiper}) is a sensible
intermediate step toward the process of vacancy creation in SCO.
This kind of change with doping in the site of the doped holes has
been seen in other cuprates, for instance in $\rm
YBa_{2}Cu_{3}O_{6+\delta}$, where the holes stay in chains in Cu and
O orbitals up to $\delta \sim 0.25$, only then doping the oxygen
states in the $\rm CuO_{2}$ planes.~\cite{2006Liang,2011Hawthorn}

To address the question of the orientation of the state behind SHP,
angle-resolved FY measurements have been made. A quantitative
analysis of the angular dependence of FY requires measurements on
bulk crystals cut at a series of angles with respect to the
crystallographic planes, to account for footprint and
self-absorption effects~\cite{1993Eisebitt}. Superlattice samples
cannot be grown at arbitrary angles. However, the SL footprint
effects are almost identical for MCP and SHP energies because of the
very similar scattering geometry. Self-absorption effects would have
to be very strong for the observed suppression of intensity at SHP.
That SL self-absorption effects are relatively small is supported by
the very small difference in the momentum linewidth between MCP and
SHP [Fig. 1(c), inset], which shows that the entire SL is probed at
both energies. In addition, the O edge step height is approximately
the same before the alignment in Fig. 2(b), which also suggests that
self-absorption effects are relatively small at the O edge.

The MCP and SHP intensities follow a similar angular dependence
[Fig. 2(b)]. We assume that the SHP state can be associated with a
single orbital (if localized) or a collection of orbitals of the
same type (if delocalized, as for the ZRS state, corresponding to
MCP), in which case the SHP state would correspond to a band of a
specific symmetry. The similar variation with angle of MCP and SHP
intensity in angle-resolved FY measurements [Fig. 2(b)] suggests
that the state corresponding to the SHP energy is oriented as the
in-plane ZRS. Also, the MCP and SHP scattering amplitudes remain
similar over a wide angular range (Fig. 4), which would be difficult
to explain if they were polarized in different directions.
Therefore, our data provide evidence that SHP is a state at the
apical oxygen site polarized in the $a-b$ plane. There are three
relevant apical oxygen orbitals: $p_z$ pointing to the Cu in the
neighboring $\rm CuO_{2}$ plane, $p_{x,y}$ pointing to Sr or La in
the same LaO plane and $p_z$ pointing to the Sr or La in the
neighboring LaO plane. The only possibility consistent with this
interpretation is the orbital pointing to the La atoms in the same
LaO plane.

\subsection{Fluorescence yield and scattering at the La and Cu edges}

RSXS can measure
bulk~\cite{2002Abbamonte,2004Abbamonte,2005Abbamonte} and
SL~\cite{2007S,2009S,S-LNOLCO} charge order. SL scattering
measurements were made with $\pi$-polarized light at beamline X1B in
an UHV diffractometer. Before the more complex SL reflectivity at
the O edge, we present the measurements at the La and Cu edges,
which will be used to illustrate the model of Sec. III A.

FY measurements at the La $\rm M_5$ edge for SL-A and SL-C are shown
in Fig. 3(a). The scattering contrast between the SL layers from
$L=1$ to $L=7$, $\delta f^{(\rm La)}_{SL}=(f_{SCO}-f_{LCO})\bigg
|_{\rm La~edge}$, is given mainly by the difference between the
number of La atoms in the LCO and SCO layers, $\langle
t_{l,LCO}^{La} \rangle$ and $\langle t_{l,SCO}^{La} \rangle$.
Consistent with this observation, there is a large non-resonant tail
in scattering at the La edge for SL peaks, which was used to
characterize the SL structure [Fig. 1(c)]. The line shape in Fig.
3(a) is approximately the same for different $L$ because of the
relatively small contribution from the surface reflectivity compared
to the SL reflection (Sec. III A).

FY measurements at the Cu edge [Fig. 3(b)] are very similar for SL-A
and SL-C; the scattering contrast at this edge is given by a small
difference in the Cu valence, $\delta f^{(\rm
Cu)}_{SL}=(f_{SCO}-f_{LCO})\bigg |_{\rm Cu~edge}$, the difference in
the dispersion corrections at the Cu edge between the SCO and LCO
layers. Even for the SL higher doping levels, holes do not appear to
empty unusual Cu states, unlike the case of
YBCO.~\cite{2011Hawthorn} Small SL imperfections shift the peaks
from integer values. The line shape changes with $L$ from a dip on
resonance (low $L$) to a peak (high $L$). This change is not due to
absorption (the calculated absorption depth is much larger than the
SL thickness), refraction (from two-dimensional profiles, data not
shown) or multiple-scattering at low $L$ (similar effects are seen
at high $L$ in manganite SL, data not shown).

\begin{figure}
\centering\rotatebox{0}{\includegraphics[scale=0.7]{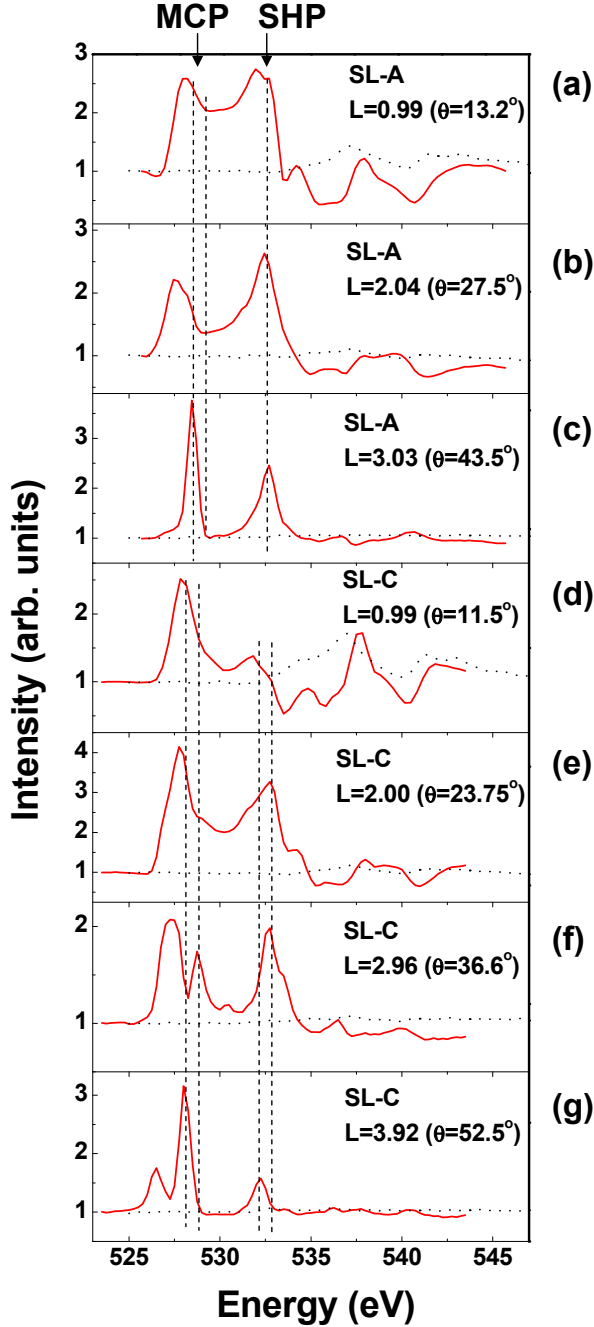}}
\caption{\label{fig:Figure5} (Color online) Line shapes at constant
$L$ for different SL peaks (solid line), compared to substrate
reflectivity (dotted line) at the O edge, for SL-A (a-c) and SL-C
(d-g). Results on SL-B (not shown) were similar to those for SL-A,
supporting the observation that small structural irregularities do
not affect our conclusions. The scans have been normalized to unity
below the edge. SL-A and SL-C have different scattering line shapes
at high $L$ because of different SL roughness and structure (Sec.
III B). Measurements on the substrate, approximating the surface
reflectivity $S_0$ (Sec. III A), show that the surface reflectivity
resonant contribution gets negligible at higher $L$, where the
substrate reflectivity becomes more featureless. Vertical dashed
lines are guides to the eye.}
\end{figure}

\subsection{Scattering at the O edge}

RSXS measurements at the O edge (Figs. 4 and 5) probe the spatial
distribution of holes in the SL. Consistent with FY measurements
(Sec. II B), the absence of a scattering peak at UHB shows that the
UHB states have been uniformly removed in all layers. The scattering
contrast in Fig. 4 occurs mostly at the two energies (MCP and SHP),
where SL-A and SL-C differed in the FY (Fig. 2). Because scattering
peaks at SL reflections (Fig. 5), the charge density is modulated at
MCP and SHP energies between SCO and LCO layers of the SL.

Sharpening of the features is observed at higher $L$ (Fig. 4). The
MCP and SHP peaks are well-separated (the additional splitting of
the MCP in SL-C will be discussed in Sec. III B). It is difficult to
explain this clear separation with a difference in the energy of the
same MCP hole state in the LCO and SCO layers, given the inherent
interface roughness of the structure. This is strong evidence for
considering SHP a qualitatively distinct hole state.

The low L line shapes for SL-A, SL-B (not shown) and SL-C are more
similar than the high L line shapes because of the reduced
importance of roughness (Sec. III B). At low L, the energy profiles
show increased intensity between MCP and SHP [Figs. 4 and 5(a)],
unlike the La $\rm M_5$ and $\rm M_4$ edges (Fig. 3). Although the O
edge line shapes are more complex, they can be qualitatively
analyzed with the same model as for the La and Cu edges (Sec. III
A). The difference in the MCP and SHP line shapes at low $L$ is due
to different interference conditions with the surface reflection
(Sec. III B).

Vacancies are present~\cite{2009Geballe} in bulk SCO as well as at
certain interfaces, even when the bulk materials do not contain
vacancies.~\cite{2006Nakagawa} If there were vacancies in the SCO
layers or at interfaces, these would give a peak in scattering at
all O edge energies. However, we do not observe large features above
the SHP energy that would indicate a structural difference (either
vacancies or oxygen atoms at interstitial sites), between the LCO
and SCO layers (Fig. 5). Since the samples were annealed in ozone,
the SCO layer thickness is relatively small, and there is little SL
scattering at the O edge other than at MCP and SHP (Fig. 5), we will
neglect vacancies in the following analysis, that is $\delta \approx
0$. The number of oxygen atoms is approximately the same throughout
the SL, while their valence is different in the SCO and LCO layers,
giving the scattering contrast. Therefore, SL scattering at the MCP
and SHP energies probes the difference between SCO and LCO layers
form factors $\delta f^{(\rm MCP, SHP)}_{SL}=(f_{SCO}-f_{LCO})\bigg
| _{\omega=\rm MCP, SHP}$, measuring the change in the dispersion
corrections with doping.

\section{Discussion}

\subsection{Scattering model}

To analyze the line shapes at the O edge, we develop a model for the
interference between the surface and a SL reflection. The X-ray
scattering intensity is $I=A|S|^2$ , with the structure factor given
by:
\begin{eqnarray}
S(\omega, Q)=\sum_{l,n}{f_{n}(\omega, Q) t^{n}_{l} e^{iQz_{l}}}
\end{eqnarray}
\noindent where $\omega$ is the incident X-ray energy, $Q=(0,0,2\pi
L/c_{SL})$ is the scattering momentum in the reflectivity geometry,
and $t_{l}^{n}$ is the occupation factor in atomic plane $l$ of
element and valence $n$.~\cite{S-LNOLCO} The form factor $f(\omega,
Q)$ for soft X-ray momenta is $f(\omega,
Q)=f^{0}(Q)+f'(\omega)+if''(\omega) \approx
f^{0}(0)+f'(\omega)+if''(\omega)$. For only one source of scatterers
($n=1$), the variables $\omega$ and $Q$ in the structure factor
$S(\omega, Q)$ are separable, that is $S(\omega, Q)=g(\omega)h(Q)$.
This implies identical line shapes (up to an overall scaling factor)
for the same edge at all $L$. This is not observed at the Cu edge or
for the MCP state [Figs. 3(a) and 4]. Therefore, at least two
scattering sources, interfering in the total structure factor
$S(\omega, L)$, need to be considered. ``Stray light" only
contributes an overall background level. To interfere, the
contributions to $S(\omega, L)$ must be coherent and of the same
energy.

With the above approximation for $f(\omega, Q)$, the structure
factor for more than one type of scatterer (the condition of a SL)
becomes:
\begin{eqnarray}
S^{ij}(\omega, L)=\sum_{n}f_{n}^{ij}(\omega) \sum_{l}\langle
t^{n}_{l}\rangle e^{2\pi iLz_{l}/c_{SL}}
\end{eqnarray}
\noindent where $\rho^{n}(L)=\sum_{l}\langle t^{n}_{l}\rangle
e^{2\pi iLz_{l}/c_{SL}}$ is the Fourier transform of the
distribution $\{ \langle t^{n}_{l}\rangle \}$ of in-plane averages
of the occupation factors for layer $l$ and element and valence $n$.
Considering only two elements, $n=\rm A$ and $n=\rm B$ as an
example, and $\langle t^{\rm A}_{l}\rangle+\langle t^{\rm
B}_{l}\rangle=1$ for all $l$, we obtain a total structure factor
$S^{ij}(\omega, L)=S_{0}^{ij}+S_{SL}^{ij}$ made of two terms:
\begin{eqnarray}
S_{0}^{ij}=\bigg( f_{\rm A}^{ij}\frac{N_{\rm A}}{N}+f_{\rm
B}^{ij}\frac{N_{\rm B}}{N} \bigg
)\rho_{0,SL}=f_{0,SL}^{ij}\rho_{0,SL}
\end{eqnarray}
\noindent and
\begin{eqnarray}
S_{SL}^{ij}=(f_{\rm B}^{ij}-f_{\rm A}^{ij})\bigg [ \rho^{\rm
(B)}_{SL}\frac{N_{\rm A}}{N}-\rho^{\rm (A)}_{SL}\frac{N_{\rm B}}{N}
\bigg ] =\delta f_{SL}^{ij} \rho_{SL}
\end{eqnarray}
\noindent where $N_{\rm A}$, $N_{\rm B}$, $N=N_{\rm A}+N_{\rm B}$
are the number of ML in one A layer, one B layer and one
superperiod, $f_{0,SL}^{ij}$ is the average of the SL layers form
factors, $\delta f_{SL}^{ij}=f_{\rm B}^{ij}-f_{\rm A}^{ij}$ is the
contrast between the form factors of the SL layers, and the
substrate contribution has been neglected.

\begin{figure}
\centering\rotatebox{0}{\includegraphics[scale=0.7]{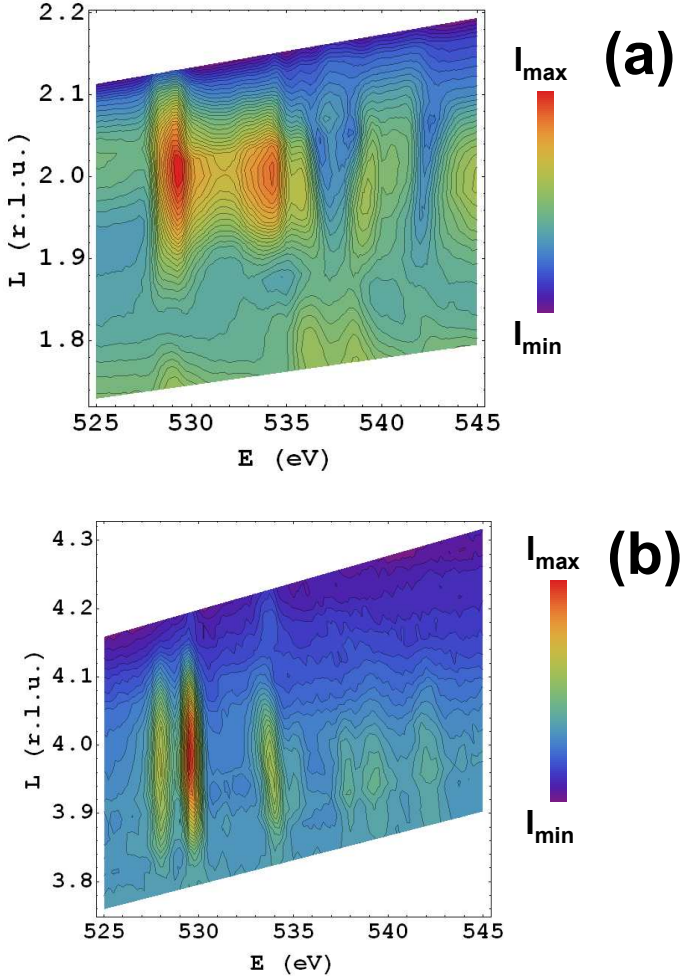}}
\caption{\label{fig:Figure6} (Color online) Two-dimensional plots in
variables energy $E$ and momentum $L$ for two selected SL peaks. (a)
Scattering two-dimensional resonance profile (logarithmic scale) for
SL-C near L=2 at the O edge shows the mirror symmetry of the MCP and
SHP scattering line shapes. Same results were obtained for SL-A and
SL-B. (b) Scattering two-dimensional resonance profile (logarithmic
scale) for SL-C near L=4 at the O edge shows that the MCP and SHP
holes are qualitatively different.}
\end{figure}

A complex oxide SL has more than two constitutive elements. We
neglect for soft X-ray momenta the small difference in $z_{l}$
between the $\rm SrO/LaO$ and $\rm CuO_{2}$ planes, or equivalently
between oxygen sites O(1) (in-plane) and O(2) (apical), within 1 ML.
In this case, $f_{0}$ becomes the average total form factor and Eqs.
3-4 can be applied by replacing $\rho^{\rm (A,B)}_{SL}$ with
$\rho^{\rm (LCO,SCO)}_{SL}$, $N_{\rm A,B}$ with $N_{\rm LCO,SCO}$,
and $f_{\rm A,B}$ with $f_{\rm LCO,SCO}$, the total form factors of
1 ML.

The indices $i$ and $j$ are dotted with the light polarization
vectors $\widehat{\epsilon}_{final}$ and
$\widehat{\epsilon}_{initial}$ as
$f=\widehat{\epsilon}^{*}_{i,final}f^{ij}\widehat{\epsilon}_{j,initial}$.
Only the in-plane $f^{xx}$ remains at the Cu edge, at MCP, as well
as at SHP from the similar angular dependence of FY (Sec. II B). In
this case, the angular dependence given by the double product
simplifies to $S \propto
\widehat{\epsilon}^{*}_{i,final}f^{ij}\widehat{\epsilon}_{j,initial}
\propto f^{xx}\rm sin^{2}(\theta)$, with an additional factor [$\rm
sin^{2}(\theta)$] that can be absorbed into the arbitrary units.

The momentum dependence of $S_{0}$ and $S_{SL}$ in Eqs. 3-4 is
contained in the functions $\rho_{0,SL}$, $\rho^{\rm (LCO)}_{SL}$
and $\rho^{\rm (SCO)}_{SL}$. For a SCO-LCO superlattice with no
roughness, they are ($r$ is the number of repeats):
\begin{eqnarray}
\rho_{0,SL}=\sum_{l, All} e^{2 \pi iLz_{l}/c_{SL}}=\frac{1-e^{2\pi i
Lr}}{1-e^{2 \pi iL/N}} \\
\rho^{\rm (LCO)}_{SL}=\sum_{l, LCO} e^{2\pi iLz_{l}/c_{SL}}
\end{eqnarray}
\noindent and
\begin{eqnarray}
\rho^{\rm (SCO)}_{SL}=\sum_{l, SCO} e^{2\pi
iLz_{l}/c_{SL}}=\rho_{0,SL}-\rho^{\rm (LCO)}_{SL}
\end{eqnarray}
\noindent Except near $L=mN$, where $m$ is an integer, the functions
$\rho^{\rm (LCO)}_{SL}$ and $\rho^{\rm (SCO)}_{SL}$ are related as
$\rho^{\rm (LCO)}_{SL} \approx -\rho^{\rm (SCO)}_{SL}$. Therefore,
they are out-of-phase:
\begin{eqnarray}
\rm {Arg}[\rho^{\rm (LCO)}_{SL}(L)]=\pi+\rm{Arg}[\rho^{\rm
(SCO)}_{SL}(L)]
\end{eqnarray}
\noindent where $\rm Arg[\rho_{SL}^{(LCO)}(L)]=\pi L \rm
(N_{LCO}-1)/N$.

The $S_{0}$ term in Eq. 3 gives the reflection from the
discontinuity in the index of refraction $n$, or equivalently in the
average form factor $f_{0,SL}$, at the sample surface. For a SCO-LCO
SL, $f_{0,SL}=\frac{N_{\rm LCO}}{N}(2f_{\rm La})+\frac{N_{\rm
SCO}}{N}(2f_{\rm Sr})+f_{\rm Cu}+ 4f_{\rm O}$, which can be
calculated from tabulated~\cite{1993Henke} values. $f_{0,SL}$ is
made mostly of non-resonant terms at the Cu and O edges (but not at
the La edge~\cite{INPREP}) and, because of this, does not depend
strongly on energy or polarization. Considering the substrate
extends the sum in Eq. 2 to an infinite number of layers, with the
function $\rho_{0,SL}$ replaced by $\rho_{0}(L)=1/(1-e^{2 \pi
iL/N})$ and $f_{0,SL}$ replaced by $f_{0}=(1-\tau) f_{0,SL}+\tau
f_{0,LSAO}$, where $f_{0,LSAO}$ is the substrate contribution and
$\tau (\theta, 2\theta, \omega)$ is a weighting factor which depends
on the scattering geometry ($\theta, 2\theta$) and energy ($\omega$)
through a variable absorption depth.

\begin{figure}
\centering\rotatebox{0}{\includegraphics[scale=0.65]{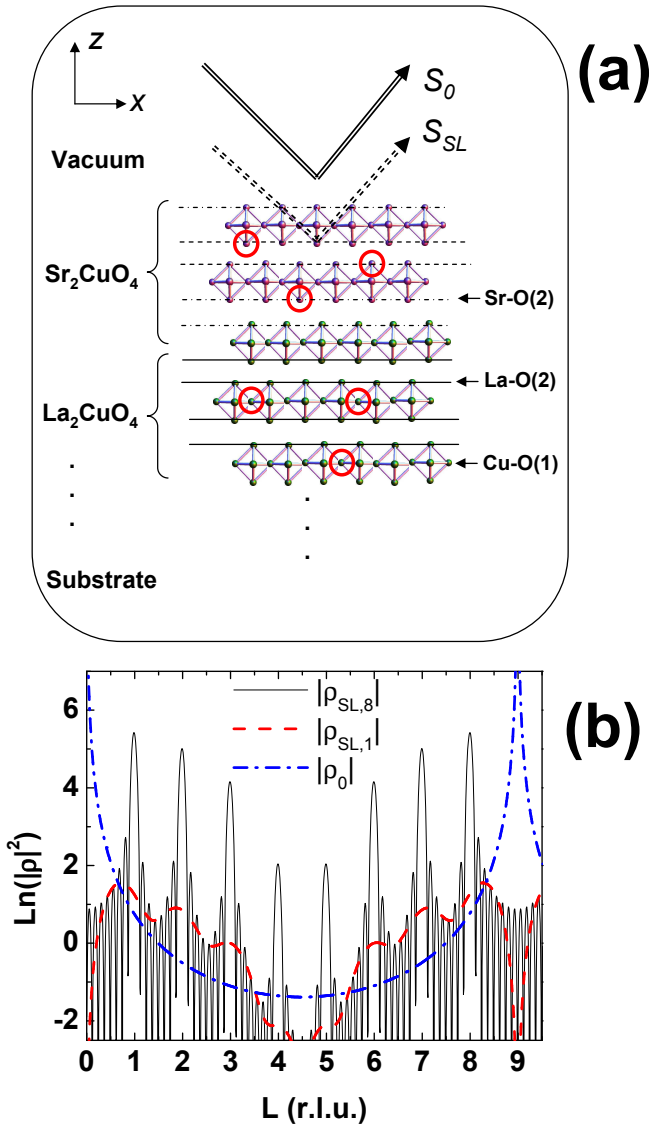}}
\caption{\label{fig:Figure6} (Color online) (a) Sketch of the two
components in the structure factor $S$: the surface $S_{0}$, from
the discontinuity between the vacuum and the surface, and $S_{SL}$,
from the modulation $\delta f_{SL}$ between SCO and LCO layers. The
circles show the location of the MCP and SHP holes doped in the LCO
and SCO layers, if the SHP is scattering from a hole on an orbital
at the apical O(2) sites (Sec. II B). A few examples of atomic
planes in the SCO and LCO layers are indicated. (b) Calculations of
$|\rho_{SL,8}(L)|$ and $|\rho_{SL,1}(L)|$ for a $\rm [2\times
SCO+7\times LCO]$ SL with no roughness and 8 and 1 repeats, compared
to the surface $|\rho_{0}(L)|$.}
\end{figure}

The $S_{SL}$ term is the reflection from the SL modulation. The
modulation of the La and Sr numbers in the SL also gives modulations
of the form factors at the Cu ($\delta f^{(\rm Cu)}_{SL}$) and O
($\delta f^{(\rm MCP,SHP)}_{SL}$) edges. Therefore, the interference
at different edges will be analyzed in terms of two main components:
a SL reflection $S_{SL}$ and a surface reflection $S_{0}$ [Fig.
6(a)]. Fig. 6(b) shows the functions $|\rho_{0}(L)|$ and
$|\rho_{SL}(L)|$ for a $\rm 8\times [2\times SCO+7\times LCO]$
structure. The $L$ dependence of $|\rho_{0}|$ and $|\rho_{SL}|$ is
very different; while $|\rho_{0}|$ has peaks at $L = m N$, where $m$
is an integer, $|\rho_{SL}|$ follows a momentum dependence with
pronounced peaks at integer $L \neq m N$. $|\rho_{SL}|$ is therefore
much higher than $|\rho_{0}|$ at these $L$ values:
$|\rho_{SL}|>>|\rho_{0}|$. It would appear that the surface
reflection can be neglected near SL reflections. However, $|f_{0}|$
is considerably larger than $|\delta f_{SL}(\omega)|$ at certain
edges, with the SL and surface structure factors comparable in
magnitude $|S_{SL}| \sim |S_{0}|$.

Including both terms, the intensity of X-ray scattering can be
written as:
\begin{eqnarray}
I=A |\rho_{SL}|^2 \Bigg |\delta
f_{SL}+\frac{f_{0}\rho_{0}}{\rho_{SL}}\Bigg|^2 =A' \big|\delta
f_{SL}(\omega)+z_{L}\big|^2
\end{eqnarray}
with $|\rho_{SL}|^2$ (independent of $\omega$) absorbed into the
arbitrary units. It is not possible at present to calculate from
first principles the energy-dependent form factor $\delta
f_{SL}(\omega)$ for a correlated oxide SL. However, FY measurements
(Fig. 2) show that the energies of MCP and SHP do not depend on the
environment of the oxygen atom, which suggests a simplified model.
Therefore, the line shapes at different edges will be modeled with a
harmonic oscillator functional dependence. The function used for the
imaginary part of the form factor is $\delta
f_{SL}''(\omega)=\frac{\alpha E}{(E-E_{0})^2+\Gamma^2}$. This is
Kramers-Kronig transformed to obtain the real part $\delta
f_{SL}'(\omega)$ [Fig. 7(a)]. Fig. 7(b) shows $|\delta
f_{SL}+z_{L}|$ for different $z_{L}$.

The interference between the surface $S_{0}$ and superlattice
$S_{SL}$ terms can qualitatively describe the scattering line shapes
at all edges. At the La edge, $|\delta f^{\rm (La)}_{SL}|$ is
comparable to $|f_{0}|$, $|S_{SL}|>>|S_{0}|$ and little change of
the resonance line shapes with $L$ is expected. Indeed, there is
little change with $L$ at the La edge [Fig. 3(a)], except at low L.
Measurements on the substrate at the La edge [Fig. 3(a)] indicate a
strong resonance in $f_{0,LSAO}$, and therefore $f_{0}$, which is
responsible for the variation in the line shape at low $L$.
Refraction effects shift the peaks and complicate the interpretation
of the measurements at low $L$; because of this, the measurements at
the La edge will be discussed in detail separately.~\cite{INPREP}
The scattering line shape at the $\rm M_5$ and $\rm M_4$ edges has
the same shape because it originates in states on the same La atoms;
this is in contrast to the scattering at the O edge at low $L$,
where the line shape at the MCP and SHP resonances is different
(Sec. III B).

At the Cu edge, $\rm \delta f^{\rm (Cu)}_{SL}$ is the change in the
dispersion corrections with doping and, because of the small
modulation in the Cu valence, $|\delta f^{\rm (Cu)}_{SL}| <<
|f_{0}|$ and $|S_{SL}|<<|S_{0}|$. Strong interference effects are
expected and observed [Fig. 3(b)]. The scattering shows a strong
surface contribution to reflectivity. The non-resonant ($f_{0,SL}$
and $f_{0,LSAO}$) form factors have large imaginary parts $f_{0}''$
because of the neighboring La edge and the line shapes resemble
those of panels (2) and (8) in Fig. 7(b).

The O edge scattering, with $|S_{SL}|\sim |S_{0}|$, is an
intermediate case. $\rm \delta f^{\rm (MCP,SHP)}_{SL}$ is relatively
large and $f_{0}''$ is smaller than at the Cu edge, with the line
shapes in a different region of Fig. 7(b): for low $L$ the MCP and
SHP are of the type (4) and (6), for high $L$ of the type (5) and
(2). Scattering line shapes are the most variable, with large
complex changes with $L$. The scattering at the O edge is analyzed
in more detail in the next section.

\begin{figure}
\centering\rotatebox{0}{\includegraphics[scale=0.5]{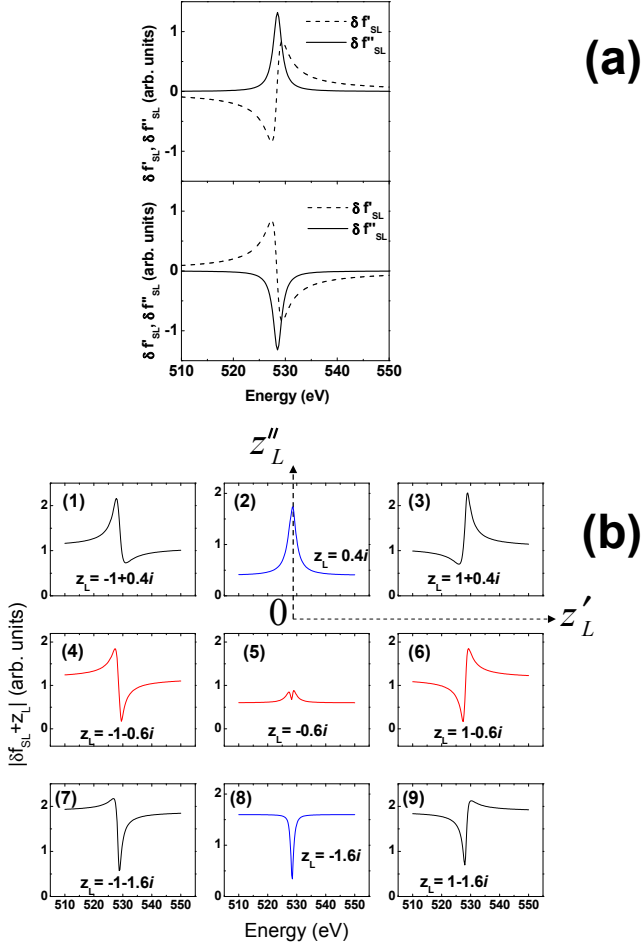}}
\caption{\label{fig:Figure6} (Color online) (a) Functions $\delta
f''_{SL}$ and $\delta f'_{SL}$ obtained by Kramers-Kronig transform.
The parameters were $\alpha=2.5 \times 10^{-3}$ (top) and
$\alpha=-2.5 \times 10^{-3}$ (bottom), $E_{0}=528.5~\rm eV$ and
$\Gamma=1~\rm eV$. (b) Line shapes $|\delta f_{SL} + z_{L}|$
calculated for different interference conditions, arranged in the
$(z_{L}',z_{L}'')$ plane according to the $z_L$ term used, show the
variety of behaviors at the Cu edge [panels (2) and (8)], La edge
[panel (4)] and O edge [panels (4) and (6)].}
\end{figure}

\subsection{Locus of MCP and SHP scatterers}

Interference between the surface and the SL reflections determines
the line shape. This can be used to find the locus of the MCP and
SHP scatterers within the SL structure.

As pointed out, the functions $\delta f_{SL}^{(\rm MCP,SHP)}$ cannot
be calculated. However, the change with doping in the dispersion
corrections at MCP and SHP energies can be estimated by looking at
the difference in FY ($\delta f''_{FY}=f''_{SL-A}-f''_{SL-C}$)
between \emph{two} differently-doped \emph{SL}, which gives the
scattering contrast between SL \emph{layers} ($\delta
f''_{SL}=f''_{SCO}-f''_{LCO}$) in \emph{one} SL (either SL-A or
SL-C). Specifically, when comparing SL-A with SL-C, FY intensity
increases concurrently at MCP and SHP [Fig. 2(a)]. Therefore, for
both MCP and SHP energies, the difference in the dispersion
corrections in the SCO and LCO layers is positive:
\begin{eqnarray}
\delta f''_{SL} ({\rm MCP,SHP}) \propto \delta f''_{FY} ({\rm
MCP,SHP}) > 0
\end{eqnarray}
\noindent The Kramers-Kronig transform of a $\delta f''_{SL}>0$ peak
gives the shape shown in Fig. 7(a) (top panel) for $|\delta f_{SL}|
= |\delta f_{SL}'+i \delta f_{SL}''|$.

However, measurements show line shapes for the MCP and SHP states
with a mirror symmetry at $L=1$ and $L=2$ (Figs. 4 and 5). It is
necessary therefore to consider the contribution to scattering of
$z_{L}$, the other factor in Eq. 9. Measured line shapes for MCP [of
type (3)-(6)-(9), right side] and SHP [of type (1)-(4)-(7), left
side] are on opposite sides of the complex $(z'_{L},z''_{L})$ plane
in Fig. 7(b). Therefore, $z^{\rm (MCP,SHP)}_L$ for MCP and SHP have
opposite phases.

Considering the factors combined in $z^{\rm (MCP,SHP)}_{L}$, since
$\rho_{0}$ is the same at MCP and SHP, and $f_0$ cannot change sign
over the $\rm \sim 4~eV$ between the MCP to SHP energies, as
evidenced by the weak energy dependence of the substrate reflection
(Fig. 4) and the large non-resonant component of $f_{0}$, the
difference between $z^{\rm (MCP,SHP)}_{L}$ must be in $\rho^{\rm
(MCP,SHP)}_{SL}(L)$, the Fourier transforms of the distributions $\{
\langle t^{\rm (MCP)}_{l}\rangle \}$ and $\{ \langle t^{\rm
(SHP)}_{l}\rangle \}$ of the occupation factors at the two energies.
Because $z_{L}$ and $\rho_{SL}$ are inversely proportional,
$\rho_{SL}$ also has opposite phases at the MCP and SHP energies:
\begin{eqnarray}
\rm{Arg}[\rho^{\rm (MCP)}_{SL}(L)]=\pi+\rm{Arg}[\rho^{\rm
(SHP)}_{SL}(L)]
\end{eqnarray}
\noindent Therefore, the distributions $\rho^{\rm
(MCP,SHP)}_{SL}(L)$ are out-of-phase, with the spatial distributions
of the MCP and SHP holes (Eq. 11) related in the same way as the
spatial distributions of the LCO and SCO layers in the SL (Eq. 8).
Then, it is necessary that the MCP and SHP hole distributions peak
in different layers.

This conclusion, using measurements at low $L$, where roughness
effects are less important, is independent of the thickness of LCO
and SCO layers in one superperiod or roughness amplitude. Roughness
effects become more important at higher $L$, where the MCP and SHP
line shapes resemble line shapes closer to the origin of the complex
plane in Fig. 7(b). We consider the $L$ dependence of the intensity
in Eq. 9, contained in the $z_{L}$ phase and amplitude. The $z_L$
phases for the MCP and SHP holes follow those of the LCO and SCO
layers, which are linear functions of $L$: $\rm
{Arg}[z^{(LCO)}_{L}]=Arg[f_{0}]+\pi L N_{SCO}/N$ and $\rm
{Arg}[z^{(SCO)}_{L}]={Arg}[z^{LCO}_{L}] +\pi$. The other
contributing factor to the change in the line shape with $L$, is the
variation in the $z_{L}$ amplitude. For Gaussian roughness, the
ratio $|\rho_{0}/\rho_{SL}|$ in $|z_{L}|$ depends on the surface
$\sigma_{s}$ and the interface $\sigma_{i}$ roughness as
$|\rho_{0}/\rho_{SL}|=|\rho_{0}/\rho_{SL}|_{\rm ideal}R(Q_{z})$,
where $R(Q_{z})=e^{-(\sigma^{2}_{s}-\sigma^{2}_{i})Q^{2}_{z}/2}$ and
$Q_{z}=2\pi L/c_{SL}$. The factor $R(Q_{z})$ becomes increasingly
important at higher $L$. Therefore, the linear increase of the
$z_{L}$ phase with $L$ and the reduction in the amplitude of
$|z_{L}|=|z_{L}|_{\rm ideal}R(Q_{z})$ at higher $L$ for
$\sigma_{s}>\sigma_{i}$ combine to make the vector $z_L$ spiral
inward in the ($z_{L}'$, $z_{L}''$) plane of Fig. 7(b) with
increasing $L$, systematically sampling different line shapes at the
same edge.

Fig. 4(c) for SL-A, with the first peak in the split MCP almost
absent, shows the interference line shape in a slightly different
location in the plane of Fig. 7(b) than the location corresponding
to the line shape in Fig. 4(g) for SL-C, where both peaks are
visible at MCP. This divergence between measurements for SL-A and
SL-C with increasing $L$ (Fig. 4) can now be explained by
considering the different roughness of SL-A and SL-C. SL-A has a
larger surface $\sigma_{s}$ than SL-C (Sec. II A) but similar
interface roughness $\sigma_{i}$. Therefore, $R(Q)$ is smaller for
SL-A, which makes the end of the $z_{L}$ vectors for SL-A and SL-C
follow slightly different trajectories with increasing $L$ in the
plane of Fig. 7(b).

The interference between $S_{SL}$ and $S_{0}$, resembling
multi-wavelength anomalous diffraction in molecular crystallography,
allows determining the superlattice MCP and SHP hole distributions
on a relatively large scale. To confirm the FY measurements of Sec.
II B and the site of the SHP state within 1 ML, a fit with the
interference model of Sec. III A would be necessary for higher wave
vectors ($L>2$), giving better spatial resolution. In future work,
fitting the evolution of measured interference line shapes (Figs. 4
and 5) with $L$ would also allow obtaining $\delta f^{\rm
(MCP,SHP)}_{SL}$, $\sigma_{s}$ and the energy-dependent $\sigma_{i}$
(on which the $z_L$ amplitude depends), and $N_{\rm SCO}$ or $N_{\rm
LCO}$ (on which the $z_L$ phase depends).

The MCP and SHP hole distributions, maximized in different layers,
are consistent with the observation that the MCP holes are
mobile~\cite{2009S} and that remotely doping a new distinct type of
holes (SHP) in the LCO layers from the Sr in the SCO layers is an
unlikely strong long-range process. In contrast, the SHP holes
remain centered on the SCO layers. The relatively smaller number of
MCP holes in the SCO layers is consistent with the saturation of FY
in Ref. 25 and observations in Ref. 32, where the apical hole site
is favored at high doping. The distributions of MCP and SHP holes
are illustrated in Fig. 6(a).

\section{Conclusion}

Using X-ray absorption and resonant soft X-ray scattering, we found
that doping in SCO-LCO superlattices empties two distinct oxygen
hole states. The location of sub-surface scatterers within the
superlattice was determined using the interference with the surface
reflection, with the distributions of the two hole states maximized
in different layers.

X-ray absorption measurements suggest that the hole state at the
higher energy is on apical oxygen atoms and polarized in the $a-b$
plane. Since the creation of a vacancy removes two holes, this would
suggest that the vacancies in bulk SCO are at the apical sites as
well.

The increase in the density of the unoccupied states at SHP has to
come from an occupied state. Photoemission experiments to determine
if a change with doping occurs in the density of states of occupied
states that mirrors the change with doping seen at SHP in this study
would be interesting.

Calculations showed that making in-plane axial orbitals more
localized, by gradually spatially removing the apical oxygen atoms,
increases $\rm T_c$,~\cite{2001Pavarini} and that, conversely,
unoccupied apical oxygen $p_z$ orbitals lower $\rm T_c$. If similar
ideas are applied to the SHP state, the doping of SHP holes might be
responsible for the decrease of $T_c$ and the hole pair breaking at
$x>x_{optimal}$~\cite{1995Emery}, and their partial removal, with
the appearance of vacancies in SCO near $x=2$, for the high $T_{c}$
of bulk SCO.

\section{Acknowledgments}

We thank A. Gozar for useful discussions. This work was supported by
the Department of Energy: RSXS measurements by grant
DE-FG02-06ER46285, superlattice growth by MA-509-MACA, NSLS
facilities by DE-AC02-98CH10886, and MRL facilities by
DE-FG02-07ER46453 and DE-FG02-07ER46471. A. R. was also supported by
MOE AcRF Tier-2 grant (MOE2010-T2-2-121) and NRF CRP.


\begin{thebibliography}{}

\bibitem{1991Eskes}
H. Eskes and G.A. Sawatzky, Physical Review B \textbf{44}, 9656
(1991).

\bibitem{1991Chen}
C.T. Chen, F. Sette, Y. Ma, M.S. Hybertsen, E.B. Stechel, W.M.C.
Foulkes, M. Schluter, S-W. Cheong, A.S. Cooper, L.W. Rupp, Jr., B.
Batlogg, Y.L. Soo, Z.H. Ming,  A. Krol, and Y.H. Kao, Physical
Review Letters \textbf{66}, 104 (1991).

\bibitem{2004Tanaka}
K. Tanaka, T. Yoshida, A. Fujimori, D.H. Lu, Z.-X. Shen, X.-J. Zhou,
H. Eisaki, Z. Hussain, S. Uchida, Y. Aiura, K. Ono, T. Sugaya, T.
Mizuno, and I. Terasaki, Physical Review B \textbf{70}, 092503
(2004).

\bibitem{1998Ronning}
F. Ronning, C. Kim, D.L. Feng, D.S. Marshall, A.G. Loeser, L.L.
Miller, J.N. Eckstein, I. Bozovic, Z.-X. Shen, Science \textbf{282},
2067 (1998).

\bibitem{1991DiCastro}
C. Di Castro, L.F. Feiner, and M. Grilli, Physical Review Letters
\textbf{66}, 3209 (1991).

\bibitem{1992Feiner}
L.F. Feiner, M. Grilli, and C. Di Castro, Physical Review B
\textbf{45}, 10647 (1992).

\bibitem{2001Pavarini}
E. Pavarini, I. Dasgupta, T. Saha-Dasgupta, O. Jepsen, and O.K.
Andersen, Physical Review Letters \textbf{87}, 047003 (2001).

\bibitem{1990deLeeuw}
D.M. deLeeuw, W.A. Groen, L.F. Feiner, and E.E. Havinga, Physica C
\textbf{166}, 133 (1990).

\bibitem{1991Ohta}
Y. Ohta, T. Tohyama, and S. Maekawa, Physical Review B \textbf{43},
2968 (1991).

\bibitem{2009Yin}
W.G. Yin and W. Ku, Physical Review B \textbf{79}, 214512 (2009).

\bibitem{1996Raimondi}
R. Raimondi, J.H. Jefferson, and L.F. Feiner, Physical Review B
\textbf{53}, 8774 (1996).

\bibitem{2006Yang}
H. Yang, Q.Q. Liu, F.Y. Li, C.Q. Jin, and R.C. Yu, Applied Physics
Letters \textbf{88}, 082502 (2006).

\bibitem{1994Al-Mamouri}
M. Al-Mamouri, P.P. Edwards, C. Greaves, and M. Slaski, Nature
\textbf{369}, 382 (1994).

\bibitem{1999Choy}
J. H. Choy, W. Lee, and S.-J. Hwang, Physica C \textbf{322}, 93
(1999).

\bibitem{1994Shimakawa}
Y. Shimakawa, J.D. Jorgensen, J.F. Mitchell, B.A. Hunter, H. Shaked,
D.G. Hinks, R.L. Hitterman, Z. Hiroi, and M. Takano, Physica C
\textbf{228}, 73 (1994).

\bibitem{1995Zhang}
H. Zhang, Y.Y. Wang, L.D. Marks, V.P. Dravid, P.D. Han, and D.A.
Payne, Physica C \textbf{255}, 257 (1995).

\bibitem{2007Yang}
H. Yang, Q.Q. Liu, F.Y. Li, C.Q. Jin, and R.C. Yu, Physica C
\textbf{467}, 59 (2007).

\bibitem{2009Geballe}
T. H. Geballe and M. Marezio, Physica C \textbf{469}, 680 (2009).

\bibitem{1989Johnston}
D.C. Johnston, Physical Review Letters \textbf{62}, 957 (1989).

\bibitem{2005Wakimoto}
S. Wakimoto, R.J. Birgeneau, A. Kagedan, H. Kim, I. Swainson, K.
Yamada, and H. Zhang, Physical Review B \textbf{72}, 064521 (2005).

\bibitem{1992Chen}
C. T. Chen, L.H. Tjeng, J. Kwo, H.L. Kao, P. Rudolf, F. Sette, and
R. M. Fleming, Physical Review Letters \textbf{68}, 2543 (1992).

\bibitem{2005Bozin}
E.S. Bozin and S.J.L. Billinge, Physical Review B \textbf{72},
174427 (2005).

\bibitem{2009Butko}
V.Y. Butko, G. Logvenov, N. Bozovic, Z. Radovic, and I. Bozovic,
Adv. Mater. \textbf{21}, 1 (2009).

\bibitem{2003Karimoto}
S. Karimoto, H. Yamamoto, H. Sato, A. Tsukada, and M. Naito, J. Low
Temperature Physics \textbf{131}, 619 (2003).

\bibitem{2009Peets}
D.C. Peets, D.G. Hawthorn, K.M. Shen, Y.-J. Kim, D.S. Ellis, H.
Zhang, S. Komiya, Y. Ando, G. A. Sawatzky, R. Liang, D.A. Bonn, and
W.N. Hardy, Physical Review Letters \textbf{103}, 087402 (2009).

\bibitem{2010Millis}
X. Wang, L. de Medici, and A.J. Millis, Physical Review B
\textbf{81}, 094522 (2010).

\bibitem{1942Hendricks}
S. Hendricks and E. Teller, Journal of Chem. Phys. \textbf{10}(3),
147 (1942).

\bibitem{1991Kuiper}
P. Kuiper, J. van Elp, G.A. Sawatzky, A. Fujimori, S. Hosoya, and
D.M. de Leeuw, Physical Review B \textbf{44}, 4570 (1991).

\bibitem{2009S}
S. Smadici, J.C.T. Lee, S. Wang, P. Abbamonte, G. Logvenov, A.
Gozar, C. Deville Cavellin, and I. Bozovic, Physical Review Letters
\textbf{102}, 107004 (2009).

\bibitem{1993Eisebitt}
S. Eisebitt, T. Boske, J-E. Rubensson, and W. Eberhardt, Physical
Review B \textbf{47}, 14103 (1993).

\bibitem{1998Kuiper}
P. Kuiper, J. van Elp, D.E. Rice, D.J. Buttrey, H.-J. Lin, and C.T.
Chen, Physical Review B \textbf{57}, 1552 (1998).

\bibitem{1988Guo}
Y. Guo, J.-M. Langlois, and W.A. Goddard III, Science \textbf{239},
896 (1988).

\bibitem{2006Liang}
R. Liang, D.A. Bonn, and W.N. Hardy, Physical Review B \textbf{73},
180505 (R) (2006).

\bibitem{2011Hawthorn}
D.G. Hawthorn, K.M. Shen, J. Geck, D.C. Peets, H. Wadati, J.
Okamoto, S.-W. Huang, D.J. Huang, H.-J. Lin, J. D. Denlinger, R.
Liang, D.A. Bonn, W.N. Hardy, and G.A. Sawatzky, Physical Review B
\textbf{84}, 075125 (2011).

\bibitem{2002Abbamonte}
P. Abbamonte, L. Venema, A. Rusydi, G.A. Sawatzky, G. Logvenov, I.
Bozovic, Science \textbf{297}, 581 (2002).

\bibitem{2004Abbamonte}
P. Abbamonte, G. Blumberg, A. Rusydi, A. Gozar, P.G. Evans, T.
Siegrist, L. Venema, H. Eisaki, E. D. Isaacs, and G.A. Sawatzky,
Nature \textbf{431}, 1078 (2004).

\bibitem{2005Abbamonte}
P. Abbamonte, A. Rusydi, S. Smadici, G.D. Gu, G.A. Sawatzky, and
D.L. Feng, Nature Physics \textbf{1}, 155-158 (2005).

\bibitem{2007S}
S. Smadici, P. Abbamonte, A. Bhattacharya, X. Zhai, B. Jiang, A.
Rusydi, J.N. Eckstein, S.D. Bader, and J-M. Zuo, Physical Review
Letters \textbf{99}, 196404 (2007).

\bibitem{S-LNOLCO}
S. Smadici, J.C.T. Lee, J. Morales, G. Logvenov, O. Pelleg, I.
Bozovic, Y. Zhu, and P. Abbamonte, Physical Review B \textbf{84},
155411 (2011).

\bibitem{INPREP}
Manuscript in preparation.

\bibitem{2006Nakagawa}
N. Nakagawa, H.Y. Hwang, and D.A. Muller, Nature Materials
\textbf{5}, 204 (2006).

\bibitem{1993Henke}
B.L. Henke, E.M. Gullikson, J.C. Davis, Atomic Data and Nuclear Data
Tables \textbf{54}, 181 (1993).

\bibitem{1995Emery}
V.J. Emery and S.A. Kivelson, Nature \textbf{374}, 434 (1995).


\end{thebibliography}
\end{document}